# Choix explicatifs chez les enseignants de physique débutants

## Quand « l'efficience mathématique » s'impose

Laurence Viennot
Université de Paris
Matière et systèmes complexes UMR 7057

**Résumé :** Des investigations ont été récemment menées sur les analyses critiques d'enseignants de physique débutants (EDs) confrontés à des explications contestables. Ces études soulèvent la question des choix effectués pour leur enseignement par des enseignants une fois qu'ils ont pris conscience d'une (ou des) faille(s) des explications analysées. Cette présentation se centrera sur les conflits possibles, chez les enseignants débutants, entre divers critères de choix déclarés pour leurs explications, notamment le caractère satisfaisant (des points de vue de la cohérence interne, de la complétude logique et de la conformité à des lois physiques admises) et la simplicité. Elle introduira la définition de « l'efficience mathématique » comme critère d'appréciation d'une explication, sur la base d'une étude dont les résultats en montrent le caractère prioritaire chez certains EDs, même au prix de la cohérence et de la simplicité. L'article se termine par les implications des questions abordées pour la recherche et la formation de maîtres.

**Mots-clés :** Analyse critique, Explications en physique, Choix explicatifs, Efficience mathématique

# Explanatory choices for beginning physics teachers

## "Mathematical efficiency" as a determining criterion

**Abstract :** Recent investigations have been carried out on critical analyses of beginning physics teachers confronted with questionable explanations. These studies raise the question of the choices made by teachers for their teaching once they have become aware of one (or more) flaw(s) in the explanations analysed. This presentation will focus on the possible conflicts, for beginning teachers, between various selection criteria declared for their explanations, including: appropriateness (from the points of view of internal coherence, logical completeness and compliance with accepted physical laws) and simplicity. It will introduce the definition of "mathematical efficiency" as a criterion for assessing an explanation, on the basis of a study whose results show that it is a priority for some EDs even at the cost of coherence and of simplicity. The article concludes with the implications of the issues addressed for research and teacher training

# Contexte de l'étude

Inspirées par les proclamations unanimes sur l'importance de l'esprit critique en science, des investigations de recherche ont été récemment menées sur les analyses critiques d'enseignants débutants (EDs) confrontés, lors d'entretiens individuels, à des explications contestables (Viennot et Décamp, 2018a). Les études menées font apparaître les principaux obstacles à la lucidité critique de ces enseignants. Elles soulèvent aussi, sans la viser explicitement, la question des choix effectués dans leur enseignement par ces enseignants une fois qu'ils ont pris conscience qu'une explication n'est pas satisfaisante du point de vue de l'un au moins des trois critères que sont la cohérence interne, la complétude logique et la conformité à des lois physiques admises (Viennot et Décamp, 2019). Il apparait ainsi que l'analyse critique d'explications, s'agissant d'enseignants, débouche naturellement sur la question de leurs décisions relativement à leur enseignement.

A cet égard, il n'est pas très risqué d'avancer que la plus ou moins grande simplicité d'une explication intervient très probablement dans leur choix pour une audience donnée. Il est donc *a priori* intéressant de voir si, et comment, les enseignants mettent en balance, notamment, « simplicité » et « caractère satisfaisant » (au sens que l'on vient de préciser). A la lumière d'études précédentes (Viennot, 2006), il apparaît qu'au moins un troisième critère s'invite dans le débat. Ce critère est relatif à l'« efficience mathématique » d'une modélisation au sens où celle-ci, via des calculs exacts, parvient à un résultat correct. Lorsque cette qualité se joint à une modélisation invalide, on peut y voir un frein important à l'analyse critique, comme on a pu l'observer lors d'études antérieures (Viennot et Décamp, 2018a). Mais étant donné que les cas d'efficiences mathématiques sur modélisation invalide antérieurement étudiés sont associés à des rituels d'enseignement, ainsi qu'à une simplification de l'explication valide, il est apparu souhaitable de tenter une nouvelle étude sur une explication n'impliquant pas ces caractères. Cette étude concerne une explication qui n'est ni habituelle ni simple, ce qui devrait déjà décourager des enseignants de la choisir. Cette explication est en outre non « satisfaisante ». Le fait que certains EDs reconnaissent cet état de fait et déclarent malgré tout choisir cette explication mathématiquement efficiente pour l'enseignement suggère de s'intéresser à ce possible critère de choix.

# Rationale, études antérieures et question de recherche

A en croire les cinq études sur le développement de l'analyse critique chez des EDs synthétisées dans Viennot et Décamp (2018a), les raisons de passivité critique dans cette population relèvent largement du domaine psychocognitif. De façon non surprenante, les habitudes enseignantes y occupent une place importante. C'est aussi le cas du biais de confirmation (Henderson *et al.* 2015), qui consiste à accepter un raisonnement au motif qu'il parvient à un résultat considéré comme correct. Un autre facteur émerge chez les EDs qui ne maîtrisent pas du tout le domaine abordé : le sentiment d'incompétence, qui les conduit à ne pas émettre la moindre critique avant d'avoir avancé dans la maîtrise du sujet, même si cela n'est nullement nécessaire pour émettre des réserves pertinentes (c'est la « *critique différée* »). Une très bonne maîtrise du sujet peut également constituer un obstacle à la critique : c'est « *l'anesthésie experte* », qui va probablement de pair avec une correction inconsciente en temps réel d'une explication non satisfaisante par celui qui en prend connaissance. Ces études visaient à éclairer – seulement - la question de la validité d'une explication, du point de vue des EDs. Mais, au fil des entretiens, des remarques spontanées des participants pointaient d'autres critères de choix – typiquement la simplicité d'une explication - à leurs yeux très pertinents pour fonder leurs décisions pédagogiques.

Il n'est pas sans intérêt, lorsqu'on s'intéresse aux choix explicatifs des EDs, d'avoir des éléments pour comprendre leurs difficultés sur le seul terrain de la critique et d'en inférer les conséquences sur le plan de leurs choix pédagogiques. Ainsi, si ce sont les habitudes qui bloquent la critique d'une explication, on peut penser que les EDs concernés hésiteront d'autant moins à l'utiliser dans leur enseignement qu'elle est communément mise en oeuvre, ceci malgré son éventuelle incohérence. Mais, pour autant, une étude abordant explicitement les choix explicatifs des EDs semble opportune. En conséquence, une des questions de recherche retenue pour l'étude décrite ici s'inscrit dans l'interrogation suivante :

*Quels critères les enseignants de physique débutants privilégient-ils dans leurs choix d'explications en physique ?*

Plus spécifiquement, la question qui est directement éclairée par cette étude est celle-ci :

*Dans quelle mesure le critère d'efficience mathématique peut-il influencer les choix explicatifs des EDs ? Peut-il s'imposer lorsque les choix qui en découlent ne permettent pas de satisfaire d'autres critères que l'on pourrait croire déterminants, comme la simplicité ou la cohérence ?*

## Une étude soulignant la prégnance de « l'efficience mathématique »

### Dispositif expérimental

Une étude récente (Viennot, 2020) s'intéresse à la fois aux difficultés des EDs en matière d'analyse critique et à leur choix pédagogique éventuel face à deux explications du même phénomène, l'une conforme à la physique acceptée et l'autre non. Ici, le choix a été fait d'éviter que l'explication contestée ne soit une explication habituelle, afin de séparer les paramètres de l'interprétation. 8 EDs en fin de formation (master MEEF2) ont participé à des entretiens individuels approfondis (45mn) à propos d'un corrigé d'exercice reproduit en Encadré 1. Par delà les détails du calcul, le point tout à fait paradoxal et contestable de ce corrigé est que, très classiquement, ce type d'exercice doit souligner la différence entre une transformation quasi-statique (suite d'états d'équilibre, selon la formulation habituelle) et une transformation non quasi-statique, pour laquelle les états intermédiaires de la transformation ne peuvent être décrits par des grandeurs intensives (pression et température). En revanche, dans le cas non quasi-statique et sous pression extérieure déclarée constante, le résultat $W_{irrev} = p_B (V_A - V_B)$ (expression du travail reçu par le gaz en détente non quasi-statique entre un volume $V_A$ et un volume $V_B$ sous pression extérieure $p_B$ constante) peut se calculer très simplement. C'est bien celui auquel parvient le calcul de l'Encadré 1, via un calcul exact. Mais celui-ci repose sur une modélisation à l'opposé de la cible conceptuelle de l'exercice : les grandeurs $p$ et $T$ y figurent comme si elles étaient définies en cours de transformation, sans parler de la relation des gaz parfaits, non valide hors équilibre. Mathématiquement « efficiente», cette solution est invalide sur le plan de la modélisation (sur le mécanisme de ce fait surprenant, voir Viennot, 2019). En tant que raison de passivité critique, cette situation a été labellisée (en anglais) « Misleading Mathathical Legitimacy » (*MML*).

Encadré 1. Un texte hautement contestable issu d'un fascicule de corrigés d'exercices pour étudiants de première année de licence.

---

Une mole de gaz parfait est située dans un cylindre fermé par un piston mobile. Elle subit une détente de la pression $p_A$ à la pression $p_B$ ($p_A > p_B$) à température T constante. Calculer le travail effectué sur ce gaz dans les deux cas suivants:

a- Détente irréversible*, sous pression extérieure constante ($p_{ext} = p_B$)

b- Détente réversible

*(a) Cas irréversible*
*Travail à T = constante* pour la détente irréversible d'une mole de gaz parfait: $W_{irrev}$

$dW = -pdV < 0$

$dW_{irrev} = -p_{ext}\,dV = -p_B\,dV$ (puisque $p_{ext} = p_B$).

Pour notre système: $pV = NRT$ et $N = 1$ mol. Donc $V = RT/p$ et $dV = -RT\,dp/p^2$

$dW_{irrev} = -p_B\,dV = RT\,p_B\,dp/p^2$

$W_{irrev} = \int_A^B RT\,p_B\,dp/p^2 = RT\,p_B\,[\frac{1}{pA} - \frac{1}{pB}] = p_B(V_A - V_B)$

*(b) Cas réversible (…)*

---

*L'enquêtrice explique que cette transforation "irréversible" est aussi non-quasistatique.

## Les entretiens et leur codage

La définition de deux arguments critiques (*C1*: les grandeurs intensives *p* et *T* ne sont pas définies en cours de transformation non quasistatique, *C2*: la relation des gaz parfaits *pV = NRT* est non valide dans ce cas), et, pour chaque argument, de trois niveaux de position critique (acceptation ferme du texte, critique ferme du texte, position mitigée) structurent le codage des transcripts d'entretien pour les aspects critiques. Relativement au choix d'une explication pour l'enseignement en première année de licence, les mêmes trois niveaux d'adhésion sont définis pour l'argument (*C3*): on peut choisir d'utiliser dans l'enseignement l'explication maintenant reconnue comme incohérente, sans critique. Les étapes du dialogue marquant un retrait en matière de position critique sur un argument donné (*C1, C2* ou *C3*) sont également repérées.

## Principaux résultats

Après 20' de réflexion dialoguée sur ce texte, seulement la moitié des participants avaient exprimé une critique ferme sur l'usage de grandeurs intensives non définie (*C1*). En effet, 3 EDs seulement exprimaient fermement la nécessité de renoncer à l'usage de la relation des gaz parfaits en cours de transformation, les autres étant parfois très sûrs de leur bon droit : « *pV = NRT*. C'est la loi des gaz parfaits, alors c'est sûr ». Et en fin de débat, après prise de conscience déclarée de ces incohérences, seulement la moitié des participants disaient fermement exclure de leur enseignement cette démonstration compliquée, inhabituelle et incohérente (*C3*). Un commentaire symptomatique illustre la prégnance éventuelle de l'efficience mathématique : « Ah oui … ça serait un bon exercice pourtant, c'est pour ça, moi je dirais, ben moi, c'est pas mal quand même pour mettre en application la théorie classique, je trouve, j'aime bien comme exercice ».

A l'issue de cette étude, il y a donc une forte présomption que ce qui vient d'être défini comme « efficience mathématique » (calcul exact, résultat correct) puisse apparaître comme une valeur en soi dans les critères de choix explicatifs des EDs, et cela même au prix de la cohérence, de la simplicité et malgré le caractère inhabituel.

Deux éléments viennent enrichir ce constant. L'un est qu'une fois exprimée une critique sur l'un des arguments *C1* et *C2*, les retraits critiques sont relativement rares (8 épisodes sur l'ensemble des entretiens). Autrement dit la critique sur ces arguments, une fois formulée par un ED, semble installée dans son analyse. L'autre est qu'à travers leurs commentaires les EDs caractérisent d'eux-mêmes (5/8) l'obstacle à la critique que représente l'efficience mathématique : "On a l'impression que / avec tous ces calculs, finalement, ben, c'est correct, c'est scientifique et si quelque chose bloque, c'est de sa faute » ou encore : « On ne voit pas ce qu'il y a derrière », « Il ne faut pas se cacher derrière les maths ».

## Discussion et perspectives

En termes d'effectifs, il s'agit d'une étude tout à fait exploratoire. Pour ce qui est de l'analyse critique défectueuse des enseignants, la valeur indicative de ses conclusions réside dans la convergence des résultats avec ceux d'études antérieures sur des thèmes différents de physique, typiquement la montgolfière supposée en situation isobare (Viennot, 2006). Pour ce qui est du caractère déterminant du critère d'efficience mathématique dans les choix pédagogiques, son intérêt est d'illustrer, à titre symptomatique, un cas extrême, où l'adoption de ce critère se fait aux dépends, notamment, de celui de simplicité, et malgré le caractère inhabituel de l'explication. De plus, celle-ci heurte de plein fouet la cible conceptuelle classiquement attribuée à l'exercice, qui est d'illustrer l'impossibilité d'analyser les étapes intermédiaires dans une transformation qui n'est pas quasi-statique.

Les commentaires recueillis confirment amplement qu'un traitement mathématique correct menant à un résultat correct peut bloquer la détection d'une modélisation défectueuse. On peut arguer qu'il est *a priori* difficile pour des enseignants débutants de discuter un document rédigé par un enseignant universitaire. Mais les participants à cette étude n'étaient pas de jeunes étudiants et ils n'ont exprimé aucun sentiment d'intimidation. Cela n'exclut pas la possibilité d'un blocage dû à un sentiment d'incompétence, comme on a pu en observer dans d'autres études [1]. Ce point mériterait d'être exploré plus avant, par exemple en variant le statut du texte proposé (texte trouvé sur la toile ou rédigé par un autre enseignant débutant). Pour leur part, les participants à cette étude soulignent très explicitement, à la fin de l'entretien, combien le traitement mathématique peut légitimer indûment une modélisation inappropriée.

On peut se demander si cette étude concerne un cas étrange sans lien avec la pratique courante de l'enseignement. Aussi surprenant que cela puisse paraître, il n'est pas rare que des rituels d'enseignement soient de ce type, associant correction mathématique et absurdité de modélisation (Viennot, 2006 ; Viennot et Décamp, 2018b).  Ceci suggère que l'exemple traité ici n'est pas totalement anecdotique. Un inventaire systématique de ce type de situation reste à faire.

Si l'on défend l'idée qu'il faudrait aider les enseignants à faire leur choix d'explication sur la base d'évaluations multicritériées préalables, alors les limites de l'efficience mathématique justifient des recherches ultérieures et une place explicite dans les formations d'enseignants. Il s'agit d'apprécier les conditions qui font parfois de l'efficience mathématique – considérée seule - le mobile de décisions pédagogiques hautement contestables. Il s'agit aussi de com-prendre comment mettre en lumière, en formation des maîtres, l'idée que les objectifs-mêmes de la formation en physique sont en cause dans ce débat.

## Bibliographie

Henderson, J. B., MacPherson, A., Osborne, J., & Wild, A. (2015). Beyond construction: Five arguments for the role and value of critique in learning science. *International Journal of Science Education*, *37*(10), 1668-1697.

Viennot L (2006). Teaching rituals, and students' intellectual satisfaction, *Physics Education* 41, 400

Viennot, L. et Décamp, N. (2018a). Activation of a critical attitude in prospective teachers: from research investigations to guidelines for teacher education. *Phys. Rev. Phys. Educ. Res*. 14, 010133  https://doi.org/10.1103/PhysRevPhysEducRes.14.010133

Viennot, L. et Décamp, N. (2018b). The transition towards critique: Discussing capillary ascension with beginning teachers.  *Eur. J. Phys. 39*, 045704 (2018).

Viennot, L. (2019). Misleading mathematical legitimacy and critical passivity: discussing the irreversible expansion of an ideal gas with beginning teacher, *European Journal of Physics* DOI: https://dx.doi.org/10.1088/1361-6404/ab1d8b

Viennot, L. et Décamp, N. (2019). *L'apprentissage de la critique Développer l'analyse critique en physique.* Les Ulis: EDP Sciences-UGA.

Viennot, L. (2020). Developing critical analysis in physics teachers : Which  directions to take ? *Phys. Educ.*.55 015008. https://dx.doi.org/10.1088/1361-6552/ab4f64